\documentclass[twocolumn, prl, aps, superscriptaddress, longbibliography, showpacs, amsmath, amssymb, floatfix]{revtex4-1}
\usepackage[colorlinks,linkcolor=blue,anchorcolor=blue,citecolor=blue,urlcolor=blue]{hyperref}
\usepackage{graphicx}
\usepackage{amssymb}
\usepackage{amsmath}
\usepackage{epsfig}
\usepackage{color}
\usepackage{mathtools}
\usepackage{bm}
\usepackage{bbm}
\usepackage{physics}
\renewcommand{\vec}[1]{\bm{#1}}
\begin{document}
\title{Unraveling the Mystery of Quantum Measurement with A New Space-Time Approach to Relativistic Quantum Mechanics}
\author{Wei Wen}
\email{wenwei@hut.edu.cn}
\affiliation{College of Science, Hunan University of Technology, Zhuzhou 412007, Hunan Province, China}

\date{\today }

\begin{abstract}
Quantum measurement is a fundamental concept in the field of quantum mechanics. The action of quantum measurement, leading the superposition state of the measured quantum system into a definite output state, not only reconciles contradictions between quantum and classical mechanics but also facilitates quantum state manipulations, including reading and resetting. Despite its significance, four fundamental issues -- randomness, instantaneousness, irreversibility, and preferred-basis -- continue to pose significant challenges to the broader application of quantum measurement and our overall understanding of quantum mechanics. In this work, we employ a new space-time approach to relativistic quantum mechanics to address these issues systematically. Our approach provides a comprehensive elucidation of the intricate connections between quantum measurement and quantum unitary evolution, as well as an in-depth analysis for the interdependence of non-local correlations and relativistic theories. We thereby reveal a more fundamental dynamical theory, beyond the traditional time-evolution equation in quantum mechanics, where the axioms of quantum measurement naturally emerge as a corollary. These findings contribute to the advancement of related fields, and our work holds potential implications for future research and applications in the realm of quantum mechanics.
\end{abstract}

\maketitle

In the development and advancement of quantum mechanics, quantum measurement has played a vital and non-negotiable role, serving both as an indispensable concept for understanding the underlying principles of quantum phenomena and as a foundation for the development of modern quantum technologies. The concept of quantum measurement is firstly brought out by Werner Heisenberg\cite{Heisenberg:1927aa} in 1927, and formulated as an axiomatic form of quantum mechanics by John von Neumann\cite{vonneumann1932mathematische} in 1930s. Early, it was instrumental in explicating the probabilistic nature of quantum systems as famously enunciated by Max Born's wave function interpretations\cite{Born:1926aa,heisenberg1930physical,Born1935,dirac1981principles}, and later used to bridge the divergency between superposition in quantum mechanics and definite outcomes in classical mechanics as the description of the Copenhagen Interpretation\cite{wheeler2014quantum,Zurek2003Decoherence,schlosshauer2007decoherence}. In recent years, quantum measurement is broadly used to establish the advancement of the quantum technology in quantum computation and quantum information\cite{Bennett:2000aa,pettaJR2005,nielsen2001quantum,wiseman2009quantum,bennik2004}, quantum imaging\cite{Devaux:16}, quantum sensing\cite{Degen2017}, precision measurement\cite{Rosi:2014aa,richardH2018}, quantum biology\cite{Gilaberte2019}, el al., showcasing its potential in advancing new research domains and catalyzing novel technologies.

Nevertheless, the physical essence of quantum measurement and its potential issues continue to present significant challenges for researchers. Despite the wave function collapse resulting from quantum measurement has the potential to produce a definite outcome faster than unitary evolution, the randomness and unpredictability of its outcomes remain an obstacle in obtaining the desired operational results. While controlling the output of quantum measurement remains a coveted technological goal, the axiomatic theory of quantum measurement hinders our ability to do so. Moreover, the wave function collapse caused by quantum measurement has been shown to be instantaneous, raising questions about potential violations of quantum mechanics and relativity. The preferred-basis problem is another issue in quantum measurement\cite{Zurek2003Decoherence,schlosshauer2007decoherence,Schlosshauer2005}. It poses the question ``How do we make a judgment about which kind of basis a wave function collapses into when we face an unperformed quantum measurement without relying on prior experiences?'' Based on the Copenhagen interpretation of the measurement axiom, no intrinsic physical mechanism for wave function collapse has been given, nor has the arising of quantum measurement issues been explained. This has led to many subsequent interpretations of quantum mechanics, including the many-worlds interpretation\cite{schlosshauer2007decoherence,Everett1957}, the quantum decoherence theory\cite{Zurek2003Decoherence,schlosshauer2007decoherence,Schlosshauer2005}, the DRM model\cite{bassi2003257}, the modal interpretation\cite{Dieks1994}. These theories offer new perspectives and interpretations of certain aspects of quantum measurement from different viewpoints, but to date, no theory has yet provided a complete and comprehensive explanation of the preferred basis, randomness, irreversibility, instantaneousness of quantum measurement.

In this work, we present a significant advancement in the field of quantum measurement theory by introducing the extended scalar Feynman path integral theory, a new space-time approach to relativistic quantum mechanics. It not only maintains compatibility with existing quantum mechanics frameworks but also naturally derives all the properties of quantum measurement mentioned above, including the Born's rule. 

The generalized form of the Feynman path integral is expressed as:
\begin{equation}
  K(\vec{r},t;\vec{r}_0,t_0)=C\sum_j W_j\exp(\frac{\mathrm{i}S_j}{\hbar}),
  \label{expr1}
\end{equation}
This formulation incorporates both path-dependent weight factor $W_j$ and path-independent weight factor $C$. The Feynman path integral emerges as a special case of this representation when $W_j\equiv 1$. Similar to the Feynman path integral, this generalized formulation enables the calculation of quantum state evolution at any time interval $\varepsilon$,
\begin{equation}
  \psi(\vec{r},t_0+\varepsilon)=\int K(\vec{r},t_0+\varepsilon;\vec{r}_0,t_0)\psi(\vec{r}_0,t_0)\mathrm{d}^n\vec{r}_0.
  \label{expr2}
\end{equation}
However, unlike in the Feynman path integral, which employs classical action to compute phase factors of propagators\cite{Feynman1948}, the action $S_j$ in Eq.~\ref{expr1} is the relativistic form $S_j=-\int_{\wp_j}(m_0c^2\sqrt{1-\dot{\vec{r}}^2/c^2}+U(\vec{r},t))\mathrm{d}t$. Here, $\wp_j$ represents the trajectory curve of the $j$-th path, and $\int_{\wp_j}$ denotes the path integral along the $j$-th path. The task of expanding the path integral faces challenges related to convergence, normalization, and compatibility. We address these challenges by extending path-dependent weight factor $W_j$ into
\begin{equation}
	W_j=\left(\frac{\mathcal{P}_j}{\mathbbm{P}_j}\right)^{n-2}\left(\frac{\Delta\tau_j}{2}\right)^{-n/2}.
\end{equation}
where, $n$ is the space dimension. Here, $\Delta\tau_j$, $\mathcal{P}_j$ and $\mathbbm{P}_j$ are all path-dependent physical quantities in relativity. $\Delta\tau$ represents the 4-D proper time along the $j$-th trajectory, while the expressions $\mathcal{P}_j$, $\mathbbm{P}_j$ pertain to the relativistic momentum $p$, and kinetic energy $T$, respectively, and can be written as
\begin{equation}
	\mathcal{P}_j=\int_{\wp_j} \sqrt{2m_0 T}\mathrm{d}t, \ \ \  \mathbbm{P}_j=\int_{\wp_j} |\vec{p}|\mathrm{d}t.
\end{equation}
The path-independent weight factor, $R(\vec{r},t_0)$, serves as a normalization element that does not contribute to the integration on the right-hand side of Eq.~\ref{expr2}. Its presence guarantees that right-hand side of Eq.~\ref{expr2} is equal to $\psi(\vec{r},t_0)$ as $\varepsilon$ approaches zero.

This advanced formulation carries substantial physical implications, effectively bridging the Feynman path integral with relativistic theory. In classical mechanics or when the speed of light, $c$, tends toward infinity, the momentum and kinetic energy exhibit a straightforward relationship: $|\vec{p}| = \sqrt{2m_0T}$. Consequently, $\mathcal{P}_j = \mathbbm{P}_j$, making $W_j=(\Delta t/2)^{-n/2}$ a path-independent quantity and causing Eq.~\ref{expr1} to transition to the classical Feynman path integral form\cite{Feynman1948}. This representation demonstrates that equal path weights occur exclusively in classical paths. In paths nearing or surpassing the speed of light, $|\vec{p}|\neq \sqrt{2m_0T}$, leading to a notable difference in weight factors for various paths and results in the manifestation of relativistic effects within the path integral. This kind of relativistic effects leads to novel physical insights beyond quantum locality as explored in the subsequent discussion.

\textit{Schr\"odinger Equation and Klein-Gordan Equation}:  
We start by taking the first time slice in the time evolution of the wave function, with the slice length $\varepsilon$ being infinitesimal in Eq.~\ref{expr2}. By letting $\sqrt{1-v^2/c^2}=-\mathrm{i}\sqrt{v^2/c^2-1}$ if $|v|>c$ and $\gamma^{-1}=\sqrt{1-(\vec{r}-\vec{r}_0)^2/c^2\varepsilon^2}$, we obtain $W=2^{n-1}(\gamma/\varepsilon)^2(\gamma+1)^{1-n/2}$ in this time slice, and then the evolution of the free particle plane wave function can be expressed as follows
\begin{align}
	\psi_p\left(\vec{r}, t_0+\varepsilon\right)&=\iint\cdots\int K\left(\vec{r}, t_0+\varepsilon ; \vec{r}_0, t_0\right) e^{\mathrm{i} \vec{p} \cdot \vec{r}_0 / \hbar} \mathrm{d}^n \vec{r}_0 \nonumber\\
	&=C f_{n}(H_p) e^{-\mathrm{i} H_p \varepsilon / \hbar} e^{\mathrm{i} \vec{p} \cdot \vec{r}}.
\end{align}
This relation holds in any dimension ({see supplemental material {\textcolor{blue}{S1-4}}}). $f_{n}(H_p)$ is a space-time independent quantity with the form $(8\mathrm{i}\pi\hbar c^2)^{n/2}(2H_p)^{-1}(m_0c^2+ H_p)^{1-\frac{n}{2}}$, where $H_p=\sqrt{m_0c^4+\vec{p}^2c^2}$ is the relativistic energy. If we let $C=1/f_n(H_p)$ in this relation, we can derive the evolution equation for the plane wave function under $H_p$ as $\psi_p\left(\vec{r}, t_0+\varepsilon\right)=e^{\mathrm{i} \vec{p} \cdot \vec{r}}$. Based on this, the evolution equation for a long time period is given by $\psi_p\left(\vec{r}, t\right)=e^{-\mathrm{i} \int_{t_0}^{t}H_p / \hbar \mathrm{d}t_1} e^{\mathrm{i} \vec{p} \cdot \vec{r}}$. Therefore, we can see that the path-independent weighting factor $C$ is a normalization factor. Its presence ensures that as $\varepsilon$ approaches zero, $\varphi(\vec{r},t_0+\varepsilon)\rightarrow \varphi(\vec{r},t_0)$ in path integral theories.

Since the wave function of any free particle can be expanded into plane waves $\psi=\sum_m a_m(t_0)e^{\mathrm{i}\vec{p}_m\cdot\vec{r}_0/\hbar}$ (for continuous plane wave expansion, the summation will become a Riemann sum), we have
\begin{align}
	\psi(\vec{r}, t)&=\iint\cdots\int K\left(\vec{r}, t ; \vec{r}_0, t_0\right) \psi(\vec{r},t_0) \mathrm{d}^n \vec{r}_0 \nonumber\\
		&=\sum_m a_m\iint\cdots\int K\left(\vec{r}, t; \vec{r}_0, t_0\right) e^{i \vec{p}_n \cdot \vec{r}} \mathrm{d}^n \vec{r}_0 \nonumber\\
		&=R(\vec{r},t_0)\sum_{m}a_mf(H_{p_m})e^{-\frac{\mathrm{i}}{\hbar} \int_{t_0}^t H_{p_m} d t_1} e^{\mathrm{i} \vec{p}_m \cdot \vec{r}} \nonumber\\
		&=R(\vec{r},t_0)f_{n}(\hat{H})e^{-\frac{\mathrm{i}}{\hbar} \int_{t_0}^t \hat{H} d t_1}\psi(\vec{r},t_0)
		\label{evol}
\end{align}
where $\hat{H}=\sqrt{m_0^2c^4+\hat{\vec{p}}^2c^2}$ is the Hamiltonian operator. To eliminate the pre-coefficient of $e^{-\frac{\mathrm{i}}{\hbar} \int_{t_0}^t \hat{H} d t_1}$, we just need to make $C=\hat{R}(\vec{r},t_0)=1/f(\hat{H})$, thereby obtaining $\psi(\vec{r},t)=e^{-\frac{\mathrm{i}}{\hbar} \int_{t_0}^t \hat{H} d t_1}\psi(\vec{r},t_0)$. The corresponding differential equation for Eq.~\ref{evol} is given by $\mathrm{i}\hbar\partial_t\psi(\vec{r},t)=\hat{H}\psi$. If a particle evolves in a space with potenial energy $U(\vec{r},t)=qV-q\vec{A}\cdot\dot{\vec{r}}$, where $V$ and $\vec{A}$ are respectively the scalar and vector components of the 4-Dimensional potential energy, we can bring $V$ and $\vec{A}$ into the Hamiltonian through gauge phase transformations (see supplemental material \textcolor{blue}{S5}). finally, we obtain a general wave function evolution equation below,
\begin{equation}
	\left(\mathrm{i} \hbar \partial_t-q V\right) \psi=\sqrt{m_0^2 c^4+(\mathrm{i} \hbar \nabla-q \vec{A})^2 c^2} \psi.
	\label{kgEq}
\end{equation}
Notable, in the low-energy regime where $\langle \hat{\vec{p}}^2c^2\rangle \ll m_0^2c^4$, the Schr\"odinger equation emerges as an approximation of Eq.~\ref{kgEq}, which can be expressed as $\mathrm{i} \hbar \partial_t\psi=(\mathrm{i} \hbar \nabla-q \vec{A})^2/2m_0 \psi+(m_0c^2+qV)\psi$. The term $m_0c^2$ is often ignored since it only contributes to a global phase of $\psi$ and does not affect the energy structure or state superposition.

In addition, the wave function that fulfills Eq.~\ref{kgEq} also satisfies the Klein-Gordon equation but with the positive probability density. In the Dirac equation, there exists a negative energy solution that is related to the antiparticle. If we introduce the negative Hamiltonian of particles, $\left(\mathrm{i} \hbar \partial_t-q V\right) \psi=\sqrt{m_0^2 c^4+(\mathrm{i} \hbar \nabla-q \vec{A})^2 c^2} \psi$, we could find that the superposition state $\psi_{\pm}=(\psi\pm\psi')/\sqrt{2}$  of positive and negative energy particle wave functions leads to negative probability density and fulfills the Klein-Gordon equation
\begin{equation}
	\left(\mathrm{i} \hbar \partial_t-q V\right)^2 \psi_{\pm}=(m_0^2 c^4+(\mathrm{i} \hbar \nabla-q \vec{A})^2 c^2) \psi_{\pm}.
\end{equation}
This explains why the probability density defined by Klein-Gordon equation is not positive, as $\psi_{\pm}$ is a superposition of wave functions corresponding to positive energy particles and negative energy particles that can annihilate each other.

It is worth noting that as $H_p$ is upgraded to the Hamiltonian operator $\hat{H}$ in Eq.~\ref{evol}, the normalization factor $C$ is also upgraded to the operator $\hat{R}$ that commutes with $\hat{H}$. Although this transformation may seem uncomfortable, as it makes $K\left(\vec{r},t;\vec{r}_0,t_0\right)$ become an operator and lose its propagator identity, it is a consequence of our path integral normalization requirements and reflects relativistic effects in path integrals at high system energies. In the low energy condition $\left\langle{\hat{p}}^2c^2\right\rangle\ll m_0^2c^4$, $\hat{R}\left(\vec{r},t_0\right)$ becomes a constant coefficient $\left({m_0}/{4\mathrm{i}\pi\hbar}\right)^{{n}/{2}}$ and makes $K\left(\vec{r},t;\vec{r}_0,t_0\right)$ reverts to the classical propagator function. The operator action of $\hat{R}$ causes the non-local correlation and leads to the wave function collapse process as described below.

\textit{The preferred-basis problem in quantum measurement}: The preferred-basis problem concerns the question of why certain basis possesses a privileged status in delineating the wave function collapse in a measurement, thereby determining the associated measurement outcome. Although the decoherence theory provides some insights into solving this problem \cite{schlosshauer2007decoherence,Schlosshauer2005,Joos2003}, in this work, we provide an new explanation for this issue based on the extended scaling path integral theory.

\begin{figure}
	\centering
	\includegraphics[width=0.49\textwidth]{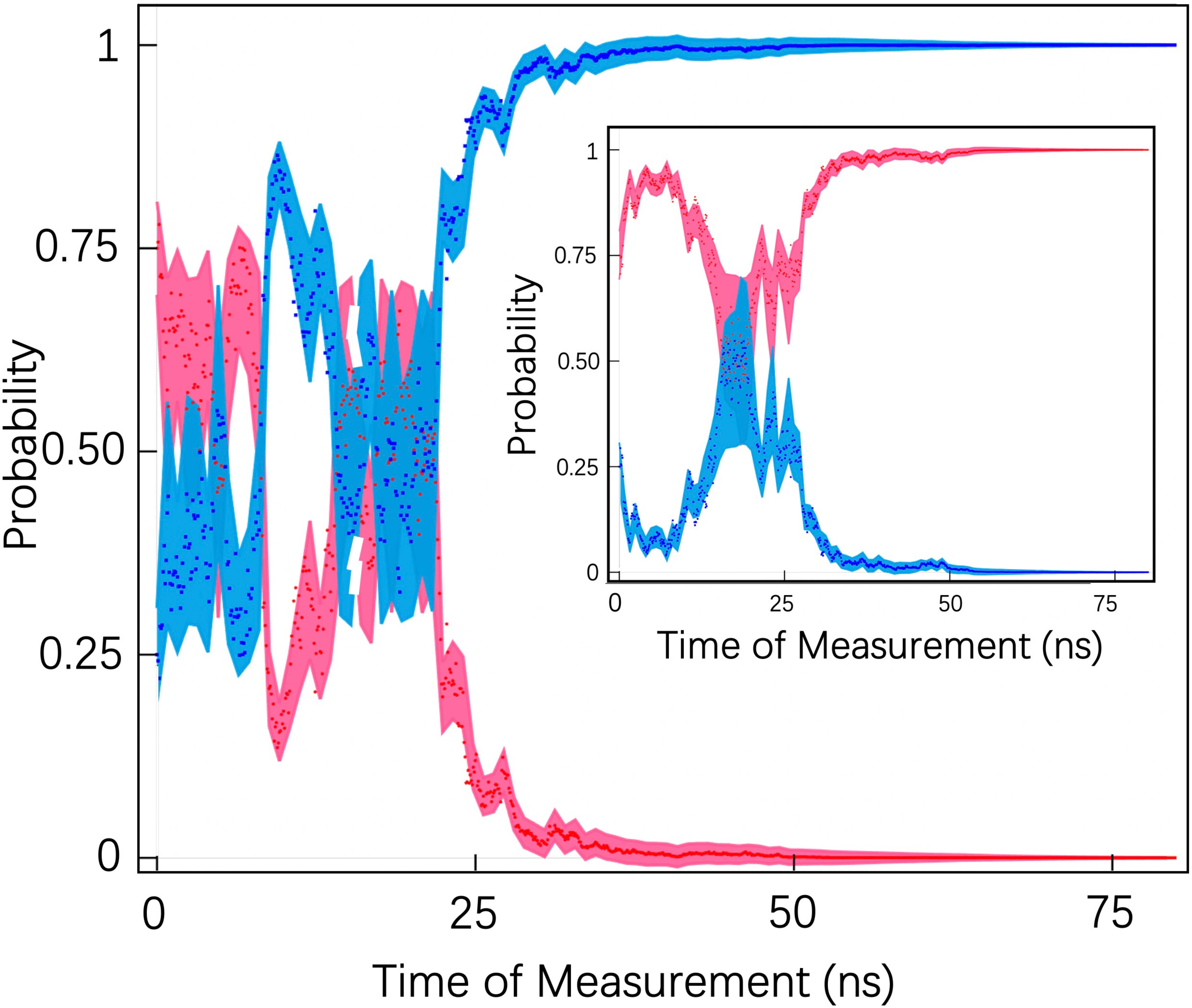}
	\caption{\textbf{Noise-induced quantum measurement and wave function collapse}. The simulated collapse process under quantum measurement for the state, $|\psi\rangle=\sqrt{3}/2|\varphi_1\rangle+1/2|\varphi_2\rangle$. The red blocks indicate the values of $|\langle \psi(t)|\varphi_1\rangle|$, while the blue blocks denote$|\langle \psi(t)|\varphi_2\rangle|$. The measuring instrument applied to the wave function generates a main potential $\hat{H}_0$, superimposed with a potential noise $\vec{A}_I$. $|\varphi_n\rangle$ are the eigenstates of $\hat{H}_0$. Due to the randomness of potential noise, $|\psi\rangle$ will randomly collapse to $|\varphi_1\rangle$ or $|\varphi_2\rangle$. In simulations, we set the environmental temperature $T$ to $300$K, the noise intensity $\sigma=42.6\mu V\cdot s/m$, and the energy gap $E_{gap}=5.0eV$.}
	\label{collapsfig}
\end{figure}

This theory suggests that quantum measurement arises due to potential energy noise within the environment. Owing to the random thermal motion of microscopic particles, macroscopic measuring instruments, while applying a main potential energy $\vec{A}_0$ to the measured quantum subsystem, generate an additional small potential energy fluctuation $\vec{A}_I$, which we term potential energy noise. Here, we take the simple and commonly used time-frequency Gaussian white noise to illustrate the wave function collapse process. For time-frequency white noise $\vec{A}_I$, the following relationships are satisfied\cite{Hollos2014}:
\begin{equation}
	\langle \vec{A}_I(t)\rangle=0; \langle \vec{A}_I(t)\cdot\vec{A}_I(t+\varepsilon)\rangle=\sigma^2\delta(\varepsilon).
\end{equation}
Given that potential energy noise is a small perturbation superimposed on the main potential energy $\vec{A}_0$, the noise intensity $\sigma^2$ is considerably smaller than $\vec{A}_0^2$. The non-differentiability of $\vec{A}$ is not important for the classical Feynman path integral since it merely contributes a random global phase to the wave function. However, for the relativisitic path integral form described by Eq.~\ref{expr1}, this non-differentiability becomes non-negligible. Based on the extended path integral theory, we can obtain the following expression
\begin{equation}
	|\psi(\vec{r},t+)\rangle=\hat{R}(\vec{r},t)\hat{R}^{-1}(\vec{r},t+)|\psi(\vec{r},t)\rangle.
\end{equation}
To conveniently represent the inner product operation of functions, we employ the Dirac notation to denote wave functions. For a time-continuous Hamiltonian, $\hat{R}(\vec{r},t)\hat{R}^{-1}(\vec{r},t+)=1$, but due to the non-differentiability of the potential energy noise, $\hat{R}(\vec{r},t)$ and $\hat{R}(\vec{r},t+)$ have a small difference, leading to (see supplemental material \textcolor{blue}{S6})
\begin{equation*}
	\hat{R}(\vec{r},t)\hat{R}^{-1}(\vec{r},t+)\simeq  1-\frac{q(\hat{\vec{p}}-q\vec{A}_0)}{m_0^2c^2\Lambda_n(\hat{H}_0)}\cdot\vec{A}_I = 1-\hat{\vec{N}}\cdot\vec{A}_I.
\end{equation*}
where $n$ is the space dimension. $\Lambda_n(\hat{H}_0)$ here is a dimensionless operator, formed by the product of $\hat{\lambda}_1=\hat{H}_0^2/(m_0^2c^4)$ and $\hat{\lambda}_2=(1+\hat{\lambda}_1)(1+n\hat{\lambda}_1/2)$. Since the operator $\hat{\vec{N}}$ commutes with $\hat{H}_0=\sqrt{m_0^2c^4+(\hat{p}-q\vec{A}_0)^2c^2}$, $\hat{\vec{N}}$ and $\hat{H}_0$ share the same set of eigenstates $\{\varphi_j\}$. Consequently, we obtain
\begin{equation}
|\psi(\vec{r}, t+)\rangle=\sum_{j=1} a_j(t)\left(1-\hat{\vec{N}} \cdot \vec{A}_I\right)\left|\varphi_j(\vec{r})\right\rangle,
\label{collpsEq}
\end{equation}
which leads to $a_j (t+)=a_j (t)(1-\vec{N}(E_j )\cdot\vec{A}_I (t+))$. Here, $\vec{N}(E_j)$ is the $j$-th eigenvalue of $\hat{\vec{N}}$. Under the influence of noise, the projection coefficients $a_j$ associated with the basis vectors $|\varphi_j\rangle$ experience continuously fluctuations with varying magnitudes due to the differing $\vec{N}(E_j)$, resulting in corresponding changes to the entire wave function $|\psi\rangle$. These changes persist in the noisy environment until $a_j=a_k\delta_{j,k}$, at which time the wave function $|\psi\rangle$ collapses into $|\varphi_k\rangle$ (see Fig.~\ref{collapsfig}). This process elucidates that the preferred basis is the eigenstate set of the Hamiltonian $\hat{H}_0$, which is determined by the main potential energy $\vec{A}_0$. These basis vectors are stably under the action of potential noise $\vec{A}_I$. 

The derivation of Eq.~\ref{collpsEq} relies on both the relativistic effects and the non-differentiability of noise. In the absence of relativistic effects, the influence of random potential energy noise would only contribute a global random phase to the wave function without affecting the probability amplitude. Moreover, for the superposition states formed by degenerate states under the time-frequency white noise, the projection coefficients of these degenerate states change synchronously due to the identical $\vec{N}(E)$ values, rendering these degenerate states inseparable and preventing the superposition state from collapsing.
\begin{figure}
	\centering
	\includegraphics[width=0.49\textwidth]{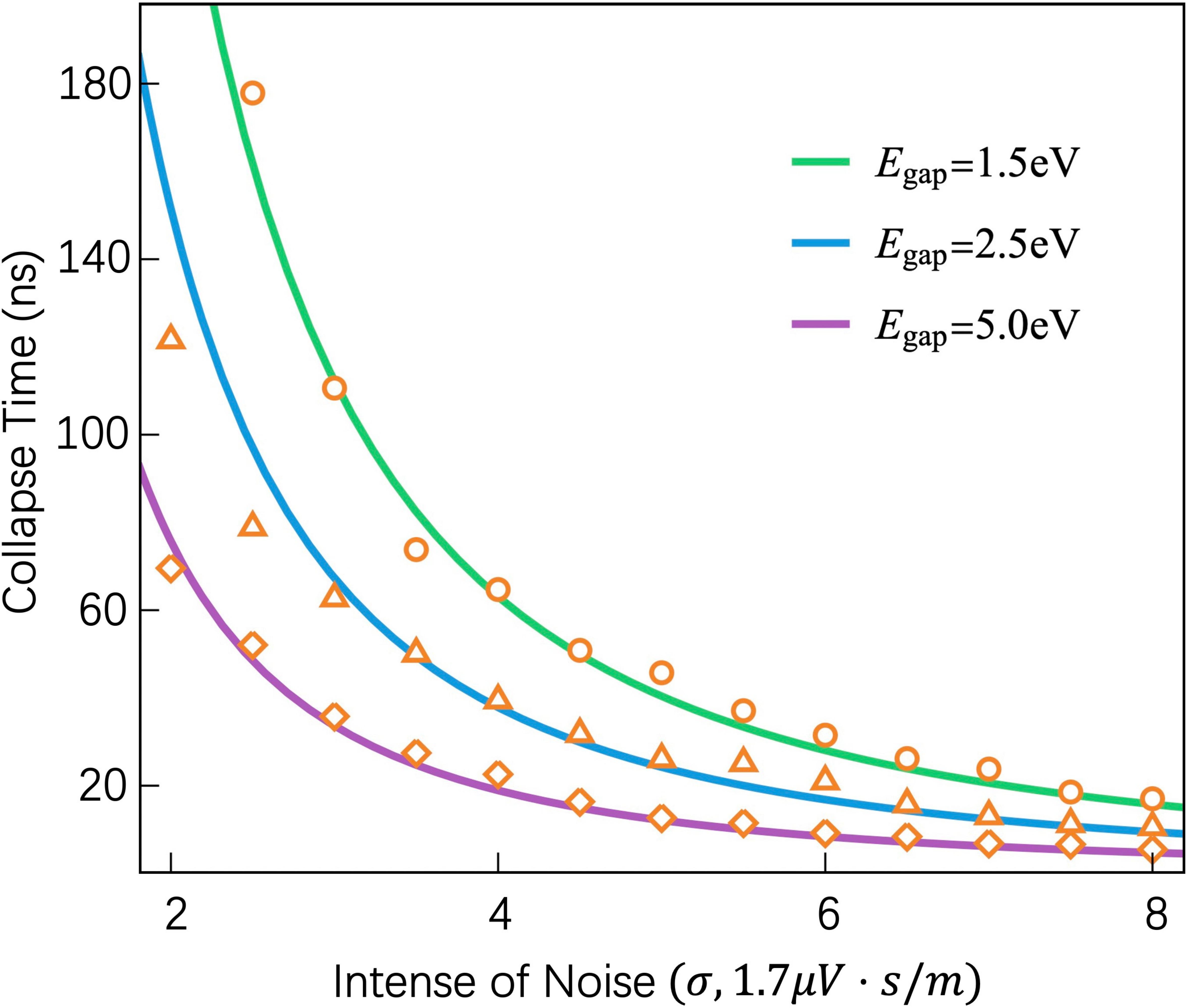}
	\caption{\textbf{Energy gap and noise influence on wave function collapse time}. We simulated the collapse process of the wave function $\psi=1/2|\varphi_1\rangle+\sqrt{3}/2|\varphi_2\rangle$ in a two-level system. Numerical simulations show that the collapse time of the wave function is influenced by the energy gap $E_{gap}$ and the potential noise intensity $\sigma$. The collapse time is proportional to $\tau_{clps}$ for Eq.~\ref{clps-time}, scaled by a factor $c_0$. Through fitting, we determined that $c_0=5.217$. Each data point is obtained through 1000 simulations.}
	\label{collapstimefig}
\end{figure}

\textit{Instantaneous Effects in Quantum Measurements}: The probability amplitude of non-degenerate states will fluctuate under the influence of potential, eventually becoming either 0 or 1, as described by Eq.~\ref{collpsEq}, thereby completing the collapse process. For a two-level system, the collapse time required can be estimated using Gambler's Ruin model\cite{grimmett2001probability,ethier2002gamblers}, given by (see supplemental material \textcolor{blue}{S7})
\begin{equation}
	\tau_{clps}\sim \frac{m_0^3c^4}{8(q\sigma)^2E_{gap}}\varepsilon,
	\label{clps-time}
\end{equation}
where $\varepsilon$ is the interval between two fluctuations of noise and $a_j(0)$ is the initial probability amplitude of the $j$-th eigenstate. the interval $\varepsilon$ depends on temperature $T$ and can be estimated using the expression $\varepsilon\sim h/(N_0^2k_BT)\approx1.03\times 10^{-15}/T$[SI](see supplemental material \textcolor{blue}{S8}). We explore the instantaneous effects in quantum measurements by simulating the collapse process of the wave function under the influence of noise, as described by Fig.~\ref{collapsfig}. Our simulations reveal that in a noisy environment, the collapse time of the wave function is extremely brief. Specifically, at room temperature, the collapse time of a two-level system is on the order of $10^{0}$ to $10^2$ nanoseconds, depending on noise intensity and magnitude of energy gap (see Fig.~\ref{collapstimefig}). This instantaneous behavior can be attributed to the superluminal effect of superluminal paths. 

In path integral calculations, superluminal paths are included to guarantee the completeness of propagators. It is these superluminal paths play a pivotal role in the short-time impulsive evolution of the wave function by connecting all parts of wave function to form a unified space-time entity. It enables the effect of any local potential applied to the wave function to propagate instantaneously throughout the entire wave function. This mechanism underlies the non-local nature of quantum measurements and constitutes the primary cause of the observed instantaneous effects. 

Given that the superluminal effect is beyond the scope of relativistic mechanics, a relevant question arises: is it feasible to exclude superluminal paths in path integral theory? The answer is negative; excluding superluminal paths in either the classical or the extended Feynman path integral would result in the failure to obtain valid differential evolution equations\cite{Redmount1993,Padmanabhan:1994aa}. Moreover, the superluminality discussed here does not imply that particles genuinely move faster than light; rather, it refers to a superluminal manifestation in the path integral form.

\textit{Randomness, irreversibility, and Born's Rule in quantum measurement}: Born's rule is a fundamental principle in quantum mechanics, stating that the probability of observing a particular state of a quantum system is equal to squared modulus of the projection coefficient of the wave function onto that state. This conclusion can be naturally obtained in our theory. Using the equation $a_j (t+)=a_j (t)(1-\vec{N}(E_j )\cdot\vec{A}_I (t+))$, we obtain $a_j^2 (t+)=(1-2\vec{N}(E_j )\cdot\vec{A}_I (t+)) a_j^2 (t)$  by neglecting higher-order infinitesimal terms. Here, $a_j^2$ is defined as $a_j a_j^{*}$ and the summation $\sum_j a_j^2$  equals 1. For long-time evolution, we have:
\begin{equation*}
a_j^2(\infty)=\lim _{\varepsilon \rightarrow 0}\left(\prod_{n=0}^{n=\infty}\left(1-2 \vec{N}\left(E_j\right) \cdot \vec{A}_I(n \varepsilon)\right)\right) a_j^2(0).
\end{equation*}
$\langle a_j^2(\infty)\rangle$ represents the statistical average of $a_j^2 (\infty)$, signifying the expected value of $a_j^2 (\infty)$ obtained after multiple wave function collapses. Form the equation above, since 
\begin{equation}
	\begin{split}
		&\left\langle \prod_{n=0}^{n=\infty}\left(1-2 \vec{N}\left(E_j\right) \cdot \vec{A}_I(n \varepsilon)\right)\right\rangle \\
	=&\prod_{n=0}^{n=\infty}\left\langle\left(1-2 \vec{N}\left(E_j\right) \cdot \vec{A}_I(n \varepsilon)\right)\right\rangle=1,
	\end{split}
	\label{averV}
\end{equation}
we have $\langle a_j^2 (\infty)\rangle=\langle a_j^2 (0)\rangle=a_j^2 (0)$. As previously discussed, under the action of potential energy noise, $a_j^2 (\infty)$ of the normalized wave function $\psi$ can only assume two values: 0 or 1. Therefore, $\langle a_j^2 (\infty)\rangle$ represents the statistical probability of $a_j^2 (\infty)=1$. The relation $\langle a_j^2 (\infty)\rangle=\langle a_j^2 (0)\rangle=a_j^{2}(0)$ tells us that the probability of the wave function collapsing into the $j$-th eigenstate during quantum measurement depends on the squared modulus of the projection coefficient of the wave function onto the $j$-th eigenstate at the beginning of the measurement. This is the Born's rule.
 \begin{figure}
	\centering
	\includegraphics[width=0.49\textwidth]{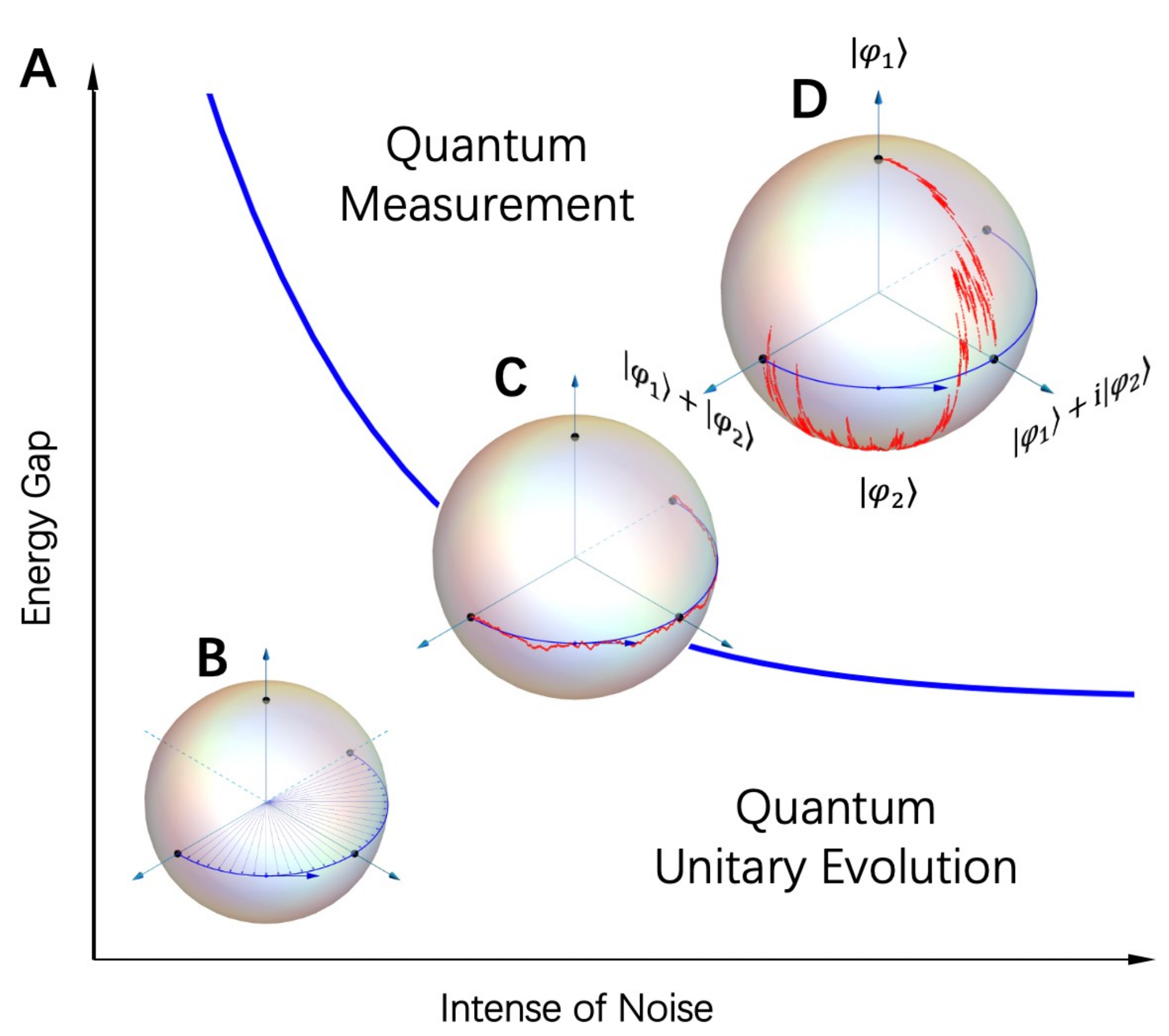}
	\caption{\textbf{Quantum measurement and quantum unitary evolution}. \textbf{A)}. Quantum measurement and quantum unitary evolution can be differentiated based on the intensity of potential noise and the magnitude of the energy gap. \textbf{B)}. For a two-level system, evolving under the main Hamiltonian $\hat{H}_0=\hbar\omega(|\varphi_1\rangle\langle \varphi_1|-|\varphi_2\rangle\langle\varphi_2|)$, the system transitions from the initial state $|\varphi_1\rangle+|\varphi_2\rangle$ to $|\varphi_1\rangle-|\varphi_2\rangle$ within half a period, $\pi/\omega$. We visualize this trajectory on the Bloch sphere, represented by the blue line. \textbf{C)}. Upon introducing weak potential noise, the trajectory of the system's wave function changes as represented by the red line. With low noise intensity, the quantum evolution prevails. The system can still evolve to $|\varphi_1\rangle-|\varphi_2\rangle$. \textbf{D)}. With the influence of strong potential noise, the wave function's change trajectory is demonstrated. Under such intense noise, quantum measurement takes precedence, leading to a swift collapse of the wave function to $\varphi_1$ within a relatively brief time interval.}
\end{figure}

\textit{Discussion and Conclusion}: Historically, Feynman's path integral theory development led to attempts to extend it into the realm of relativity, targeting a relativistic path integral for single-particle evolution. However, the non-quadratic form of relativistic action posed challenges in integration complexity, convergence, and compatibility\cite{HENNEAUX1982127,Koch:2018aa}, resulting in unsuccessful efforts by Feynman and others. Although Feynman introduced path integrals into Quantum Electrodynamics (QED) and successfully combined relativity and path integral theory from a quantum field perspective, it does not represent the single-particle relativistic path integral has been developed\cite{Koch:2018aa,Padmanabhan:1994aa}. In this work, we deviate from the field theory approach to incorporate relativistic effects into path integral theory, developing a single-particle description using relativistic path integral theory. Our findings reveal that traditional propagators used to describe quantum system evolution lead to information loss, suggesting that applying this new form of path integral to QED theory may provide modifications to propagator-based QED theory, potentially addressing accuracy issues at high energies.

Our work exhibits that the process of quantum measurement is induced by both relativistic effects and potential noise. Potential noise refers to minute fluctuations in potential energy caused by the measuring instrument on the microscopic quantum system, rendering the wave function unstable. The transition of the wave function from an unstable state to a stable state constitutes the process of measurement collapse, which, although rapid, is not ``no time spent'' as previously thought (see Fig.~1). We present simulations of quantum measurement processes, revealing that the collapse time depends on the energy gap between quantum states, the temperature of the external environment, and the intensity of fluctuations, with collapse times being on the order of $10^0$ ns to $10^2$ ns.

Furthermore, we also demonstrates that quantum measurement and quantum unitary evolution share a common origin, both stemming from path integral Eq.~\ref{expr2}. Quantum unitary evolution is determined by the main potential $\vec{A}_0$, while quantum measurement is governed by potential noise $\vec{A}_I$. When potential noise is minimal, quantum unitary transformation dominates (Fig.~2B-C); when potential noise is significant, quantum measurement dominates (Fig.~2D). To achieve wave function collapse, a sufficiently ``hot'' potential fluctuation must emerge through non-local correlations in the system, and the measurement potential generated by the measuring instrument must be ``large'' enough to distinguish between different quantum states; otherwise, unitary evolution occurs. We illustrate the boundary between these two phenomena in Fig.~2A, providing criteria to distinguish between microscopic and macroscopic systems.

Finally, we show that the probability of wave function collapse onto the eigenstates of ${\hat{\vec{H}}}_0$ can be rigorously proven to follow Born's rule according to this theory. Our theory posits that Born's rule is a physical principle that holds under white noise conditions, with different forms of noise affecting the collapse probability. If the noise is controllable and predictable, then the collapse outcomes will also be controllable and predictable.

Despite the interesting conclusions presented in this work, some detail-oriented issues warrant further discussion. In the discussion of the preferred basis, it appears that we only considered the effects of the vector potential $\vec{A}$ while neglecting the influence of the scalar potential $V$. However, the scalar potential $V$ can be integrated into the vector potential $\vec{A}$ as the Weyl gauge be adopted. Additionally, the relativistic effects discussed here refer to the paths in the path integral, not the particles themselves. Therefore, these effects persist even at low particle energies and the wave function collapse of low-energy particles can still be described by this theory. Moreover, while this article focuses on time-frequency white noise, other noise forms might affect wave function collapse time and cause collapse in superpositions of degenerate states. Further research into the effects of various noise forms on quantum measurements is warranted.

\textit{Acknowledgements}: This work is supported by Supported by National Natural Science Foundation of China (Grant No. 11904099), Natural Science Foundation of Hunan Province of China (Grant No. 2021JJ30210), and Excellent Youth Program of Hunan Provincial Department of Education (Grant No. 22B0609). 
\bibliography{ref}

\end{document}